\pdfoutput=1 
\documentclass[a4paper,fleqn]{cas-dc}

\usepackage{natbib}

\usepackage{tabularx}
\usepackage{xurl} 
\usepackage{placeins} 
\usepackage{listings}
\lstdefinestyle{sigmavql}{
  basicstyle=\ttfamily\scriptsize,
  columns=fullflexible,
  keepspaces=true,
  breaklines=true,
  breakatwhitespace=false,   
  postbreak=\mbox{\textcolor{gray}{$\hookrightarrow$}\space}, 
  frame=single,
  framerule=0.3pt,
  rulecolor=\color{black!20},
  showstringspaces=false,
  upquote=true,
  tabsize=2,
  aboveskip=0.25\baselineskip,
  belowskip=0.25\baselineskip,
  xleftmargin=0pt,
  xrightmargin=0pt,
  linewidth=\columnwidth     
}
\begin{document}
\let\WriteBookmarks\relax

\renewcommand{\topfraction}{0.9}
\renewcommand{\bottomfraction}{0.9}
\renewcommand{\textfraction}{0.07}
\renewcommand{\floatpagefraction}{0.85}

\shorttitle{Velociraptor Unified Detection-Forensics Methodology}
\shortauthors{Raesa et al.}

\title [mode = title]{Extending Detection Engineering to Digital Forensics: The Velociraptor Unified Detection-Forensics Methodology}

\author[1]{Aghni Anugrah Raesa}
\cormark[1]
\ead{a.raesa@student.uq.edu.au}

\author[1]{Adithyan Shaji Nambiar}
\ead{adithyanshajinambiar@student.uq.edu.au}

\author[1]{Veda Dawoonauth}
\ead{v.dawoonauth@student.uq.edu.au}

\author[1]{Aditya Kumar}
\ead{aditya.kumar@student.uq.edu.au}

\author[2]{Mike Cohen}
\ead{mike@velocidex.com}

\author[1]{Priyanka Singh}
\ead{priyanka.singh@uq.edu.au}

\affiliation[1]{organization={School of Electrical Engineering and Computer Science, The University of Queensland},
    addressline={Building 78 General Purpose South, Staff House Road},
    city={St Lucia},
    postcode={4072},
    state={QLD},
    country={Australia}}

\affiliation[2]{organization={Velocidex Enterprises},
    city={Gold Coast},
    state={QLD},
    country={Australia}}

\cortext[1]{Corresponding author}

\begin{abstract}
Detection engineering and digital forensics have evolved in parallel rather than
in partnership, leaving a gap between real-time alerting and forensic analysis.
This paper develops a unified detection-forensics methodology using
Velociraptor, where detection logic directly initiates targeted evidence
acquisition at the point of detection.

The contribution is threefold: (1) a four-stage methodology (baseline
establishment, evidence correlation, attack chain analysis, and scenario
labelling with confidence) that converts artefact knowledge into reusable and
testable detection rules suitable for both post-incident triage and live
monitoring; (2) a practical demonstration, using three Velociraptor BaseVQL log
sources (\texttt{forensics/windows/prefetch}, \texttt{forensics/windows/usn},
and \texttt{/windows/wmi}) that practitioners can deploy today, showing that
artefact-based detections enable scalable forensic triage without full
disk acquisition; and (3) evidence that periodic artefact analysis offers
continuous monitoring while substantially reducing data volume compared to
conventional endpoint logging.

Two case studies illustrate the approach: a Prefetch/USN baseline for triage when
Windows Event Logs are cleared or unavailable, and a WMI persistence correlation
supporting both triage and continuous monitoring through periodic artefact
analysis.
\end{abstract}



\maketitle

\section{Introduction}
\label{sec:introduction}

When responding to incidents on endpoints, there are two main goals. The first
is to establish what happened in the past, including the initial compromise
vector and gather clues of how most to efficiently evict the adversary from the
network. The second goal is to detect further adversary activity of the endpoint
to prevent further compromise~\cite{paloalto_dfir_nd}. These dual goals have
traditionally employed two distinct fields: Digital Forensics is used to
reconstruct previous activity~\cite{nist_sp800_86}, while Detection Engineering
is used to detect future compromise~\cite{lange_detection_2023}.

Digital forensics uses artifacts left behind on the endpoint to deduce past
activity. This utilises disk or memory artifacts, such as USN Journal, NTFS
timestamps, application SQLite databases to build timelines of past
activity~\cite{nist_sp800_86,shaaban_practical_2016}.

Detection systems primarily uses a forward and match model, where telemetry,
such as Windows Event Longs or Sysmon events, are forwarded to a central
Security Information and Event Management (SIEM)
platform~\cite{myllyla_detecting_2021}. The SIEM aggregates these forwarded
event logs and applies various detection rules centrally. Detection systems are
designed to handle large event volumes from many endpoints, giving wide coverage
of the monitored network~\cite{nist_sp800_92}.

\begin{figure}[pos=tbp]
    \includegraphics[width=\columnwidth]{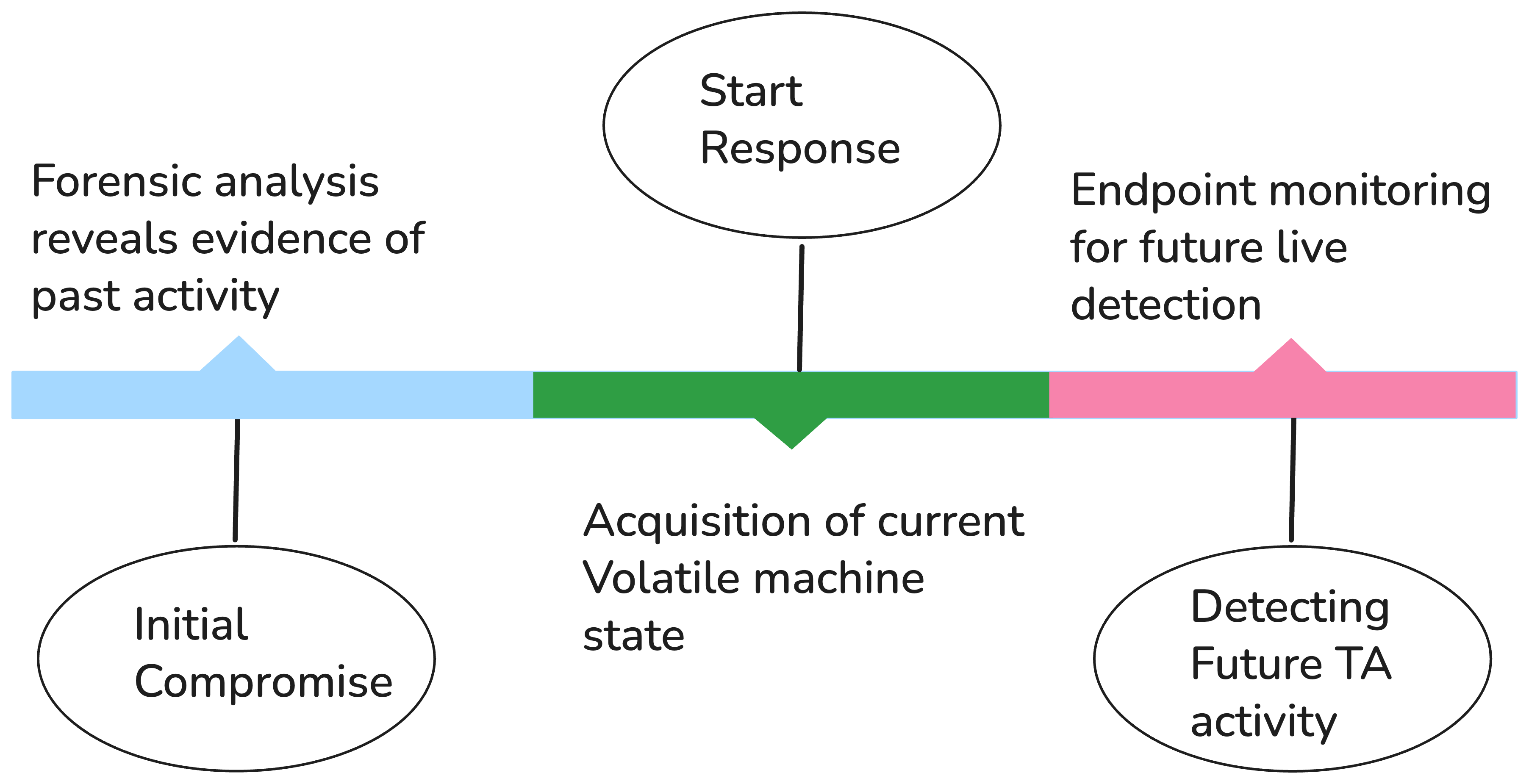}
    \caption{\textbf{The DFIR attack timeline.} Responders
      traditionally use digital forensics to reconstruct past
      activities, while using event log based live detection for
      future monitoring.}
    \label{fig:attack_timeline}
\end{figure}

This typical workflow is illustrated in Figure~\ref{fig:attack_timeline}, where
Digital Forensics provides the view into past activities, while live monitoring
provides the view into future compromise.

However, attackers routinely disable, evade, or clear these volatile log sources
(MITRE ATT\&CK T1070.001)~\cite{heiligenstein_2020_indicator}, undermining the
very data on which SIEM-based detection depends.

Another consideration in modern incident response is the speed and effectiveness
of \textbf{triage analysis}. Due to the large number of endpoints involved, a
full forensic image is rarely taken. Instead, investigators turn to forensic
triage: the rapid assessment of endpoints to identify and prioritise compromised
systems without performing full disk acquisition~\cite{shaaban_practical_2016}.
Triage relies on durable forensic artefacts, including execution traces,
filesystem journals, and persistence configurations, that persist on disk even
after event logs have been cleared~\cite{shetty_dark_2022}.

Current triage workflows remain largely manual, requiring analysts to examine
each artifact source independently across potentially hundreds of endpoints, a
process that can take days or weeks and is difficult to
standardise~\cite{ras_digital_2018}. This does not scale to large number of
endpoints.

The contrast between detection engineering and digital forensics
illustrates a persistent gap. Detection rules generate alerts, based
on a large volume of events from many endpoints, but rarely preserve
the forensic context needed for investigation. While forensic analysis
produces rich evidence but lacks the automation and scalability of
rule-driven detection. Bridging this gap requires a unified approach
in which detection logic directly initiates targeted evidence
acquisition.

This paper addresses that gap by developing a unified
detection-forensics methodology. The main innovation is moving the
detection engine to the endpoint for an endpoint-centric detection
approach.

\begin{figure*}[pos=tbp]
    \centering
    \includegraphics[width=\textwidth]{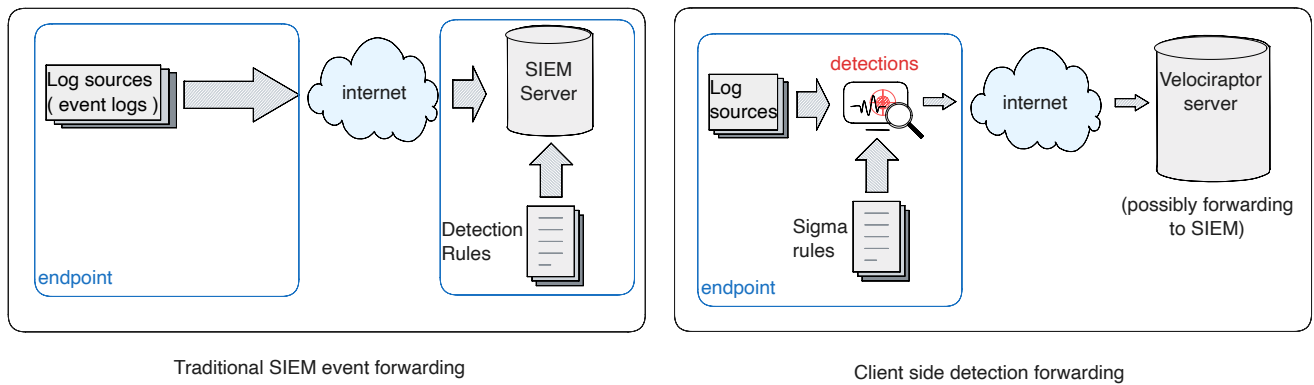}
    \caption{Traditional event forwarding detection pipelines rely on
      forwarding all events off the endpoint to a central SIEM, where
      detection rules can be applied. However, this leads to high
      event volumes and limited event types (usually only event logs
      are forwarded). However, The endpoint centric detection approach
      places detection rules on endpoint, vastly reducing the event
      volume - only high valued events are forwarded. This allows the
      use a wider range of event sources, such as forensic artifacts
      to improve detection fidelity.}
    \label{fig:velociraptor_sigma_flow}
\end{figure*}

Figure \ref{fig:velociraptor_sigma_flow} illustrates the conceptual differences
between traditionally SIEM based detection pipelines and the endpoint-centric
model. In the traditional model, event logs are forwarded with minimal filtering
from the endpoint over the network to a central SIEM - where detection rules are
applied. This typically results in large event rates from each endpoint,
transferred over the internet~\cite{nist_sp800_92}. Additionally, this model
only provides access to traditional event logs~\cite{myllyla_detecting_2021}.

On the other hand, the endpoint centric model, moves detection to the
endpoint itself, where only matching events are forwarded. This
results in a much reduced event rate, with each event of higher value
since it indicates a match with a detection rule. Additionally, by
matching the events on the endpoint itself, we may use other sources
of events to create better tuned detection rules.

The challenge is how to best develop detection rules within the
endpoint centric model, to bridge the forensic and detection gap: Speed
up forensic analysis using detection rules, and then apply these
detection rules to a live system to enhance detection efficacy.

In this work, we use Velociraptor, an open-source DFIR platform with native
Sigma rule support and VQL-based artefact collection running directly on
endpoints~\cite{velociraptor_overview_nd}. We document a methodology which can
be used to develop such forensic detection rules and demonstrate how this can be
used in two specific use cases.

The methodology treats detection and forensics as one continuous
process: every detection is designed to produce evidence-backed
results rather than alerts requiring separate
investigation.

In summary, the contributions of this paper are:
\begin{enumerate}
    \item \textbf{Methodology.} A four-stage methodology that guides
      the systematic development of forensic detections by converting
      artefact knowledge into reusable and testable detection rules,
      suitable for both post-incident triage and live monitoring.
    \item \textbf{Triage at scale.} A practical demonstration that
      artefact-based detections support scalable forensic triage,
      enabling investigators to identify and prioritise compromised
      endpoints across large environments without the need for full
      disk acquisition.
    \item \textbf{Live monitoring.} Supporting evidence that periodic
      artefact analysis offers continuous live monitoring while
      substantially reducing data volume compared to conventional
      endpoint logging, effectively connecting detection engineering
      with forensic readiness.
\end{enumerate}

To demonstrate operational usefulness, the case studies include the Sigma rule
YAML and BaseVQL log-source mappings used in Velociraptor. These are presented
as practitioner ready artifacts rather than as standalone research
contributions. This paper is organized as follows. Section~\ref{sec:background}
reviews related work,and Section~\ref{sec:methodology} presents the four-stage
methodology, illustrated through two case studies: a post-incident triage
baseline (Section~\ref{sec:resilient}) and a WMI persistence correlation
for continuous monitoring (Section~\ref{sec:confidence}).
Section~\ref{sec:discussion} discusses findings and limitations, and
Section~\ref{sec:conclusion} concludes.

\section{Background and Related Work}
\label{sec:background}
\subsection{Detection Engineering and DFIR}
Intrusion detection has long relied on monitoring system and network
events for signs of adversarial behaviour or security policy
violations \cite{bace_intrusion_2000}. Within Security Information and
Event Management (SIEM) ecosystems, detection engineering has emerged
as the practice of designing, refining, and validating rules that
surface malicious activity efficiently while balancing false positives
and false negatives
\cite{digiorgio_detection_2021,shelnt_exploring_2025}. These rules
often operate over forwarded logs such as Windows Event Logs or Sysmon
telemetry, aggregated centrally for correlation and alerting
\cite{ban_breaking_2023,yang_enhancing_2025}.

However, the growing scale and complexity of cyber incidents has
exposed inherent limitations in log-centric workflows. Attackers may
disable logging, clear event logs (e.g., MITRE ATT\&CK T1070.001)
\cite{heiligenstein_2020_indicator}, or deliberately operate in ways
that leave minimal trace in traditional log sources, thereby
undermining SIEM-based visibility \cite{mitre_attack_nd}. In parallel,
digital forensic practitioners have emphasised that investigations
increasingly involve manually reviewing large volumes of artifacts
across multiple sources, a process that can take days or weeks and is
difficult to standardise
\cite{luciano_digital_2018,soltani_detecting_2023}.

Understanding this gap requires distinguishing two operational
contexts. Live detection operates continuously over forwarded event
streams via SIEM platforms, depending on logging infrastructure
deployed before an incident \cite{sans_nextgen_siem}. Forensic triage,
by contrast, analyses historical artifacts after an incident is
suspected, reconstructing what occurred from Prefetch files, USN
journal entries, and WMI repository objects
\cite{nist_sp800_86}. Triage is critical when live logging was absent
or Windows Event Logs were cleared/disrupted (T1070.001).  Sigma rules
over Velociraptor's BaseVQL operate in this triage context, querying
durable artifacts rather than live streams
\cite{velociraptor_sigma_docs}. Case Study 1 demonstrates pure
triage-mode; Case Study 2 operates in hybrid mode.

This separation leaves a gap: detection rules raise alerts but rarely
orchestrate artifact collection, while forensic examinations proceed
without reusing detection logic. Literature calls for unified
frameworks bridging real-time detection and durable forensic evidence
\cite{hargreaves_abstract_2024,breitinger_dfrws_2024}.

\subsection{Velociraptor, Sigma, and Artifact-Centred Detection}
Open-source DFIR platforms have become central to addressing some of
these integration challenges. Velociraptor is an open-source
endpoint-focused DFIR tool that uses the Velociraptor Query Language
(VQL) to collect and analyse artifacts at scale
\cite{cohen_2025_developing}. Instead of streaming raw data to a
central store, Velociraptor executes queries on endpoints and returns
only the relevant results, supporting scalable triage and targeted
collection \cite{velociraptor_overview_nd}. Its extensive library of
artifacts covers system processes, network connections, Windows event
logs, Prefetch files, the NTFS USN journal and more.

Sigma provides a complementary capability: a vendor-agnostic rule
format that allows detections to be written once and then compiled
into queries for various backends \cite{sigmahq_about_nd}. Sigma rules
define what to detect and where to look via logsource definitions
(e.g., product, category, service) and are often mapped to MITRE
ATT\&CK tactics and techniques
\cite{mitre_attack_nd,sigmahq_about_nd}.

Recent work has explored integrating Sigma directly within the Velociraptor
ecosystem. The case study in Section~\ref{sec:resilient} demonstrates three
Sigma-compatible log sources within Velociraptor's Windows BaseVQL model for
Prefetch and USN artifacts, each immediately usable by practitioners:

\begin{itemize}
    \item \texttt{forensics/windows/prefetch} - parses Prefetch files
      to expose fields such as executable name, path, run count and
      last-run timestamps, providing durable evidence of execution.
    \item
      \texttt{process\_creation\allowbreak/windows\allowbreak/attack\_prefetch}
      - normalises Prefetch into a process-creation-like view,
      mimicking event logs even when Security/Sysmon logs are missing.
    \item \texttt{forensics/windows/usn} - normalises NTFS USN Change
      Journal entries, exposing OS path, filename, reason codes and
      timestamps for file system operations.
\end{itemize}

These log sources treat forensic artifacts as event streams, allowing Sigma
rules to operate directly on durable execution and file system evidence rather
than only on volatile logs. This supports a "forensic-aware detection
engineering" model in which the same portable rule logic can be applied across
live telemetry and post-mortem artifacts, thereby improving resilience
specifically against Windows Event Log clearing/disruption (T1070.001) and
expanding the scope of automated triage.

In parallel, the case study in Section~\ref{sec:confidence} focuses on
Windows Management Instrumentation (WMI) persistence through permanent
event subscriptions (MITRE ATT\&CK
T1546.003)\cite{murphy_2020_event}. WMI’s subscription model uses
three durable objects - Event Filters, Event Consumers and
Filter-to-Consumer Bindings, to trigger code execution based on system
conditions \cite{graeber_2015_abusing,ballenthin_2015_whymi}. These
objects are stored in the WMI/CIM repository and may execute under
WmiPrvSe.exe with SYSTEM privileges, persisting across reboots and
blending with legitimate administrative activity
\cite{marcho_2019_wmi,tilbury_2023_finding}.

To address this, the case study in Section~\ref{sec:confidence} demonstrates
deployment of Velociraptor's \texttt{/windows/wmi} log source, which enumerates
existing WMI subscriptions in a Sigma-compatible schema, exposing filter WQL,
consumer type and bindings for detection logic. Sigma rules can then flag
high-risk patterns, such as uptime-gated \_\_InstanceModificationEvent filters
tied to CommandLineEventConsumer objects and suspicious paths. The same logic
then drives targeted artifact collection, namely repository objects, Prefetch
evidence of the executed payload and SYSTEM process state confirming
execution-of-record. This “detection-as-collection” design exemplifies how
detection rules can be made operationally actionable within a forensic
methodology, rather than existing as isolated alerts.

\section{The Velociraptor Unified Detection-Forensics Methodology}
\label{sec:methodology}

\subsection{Overview and Design Goals}

\begin{figure}[pos=tbp]
    \centering
    \includegraphics[width=\columnwidth]{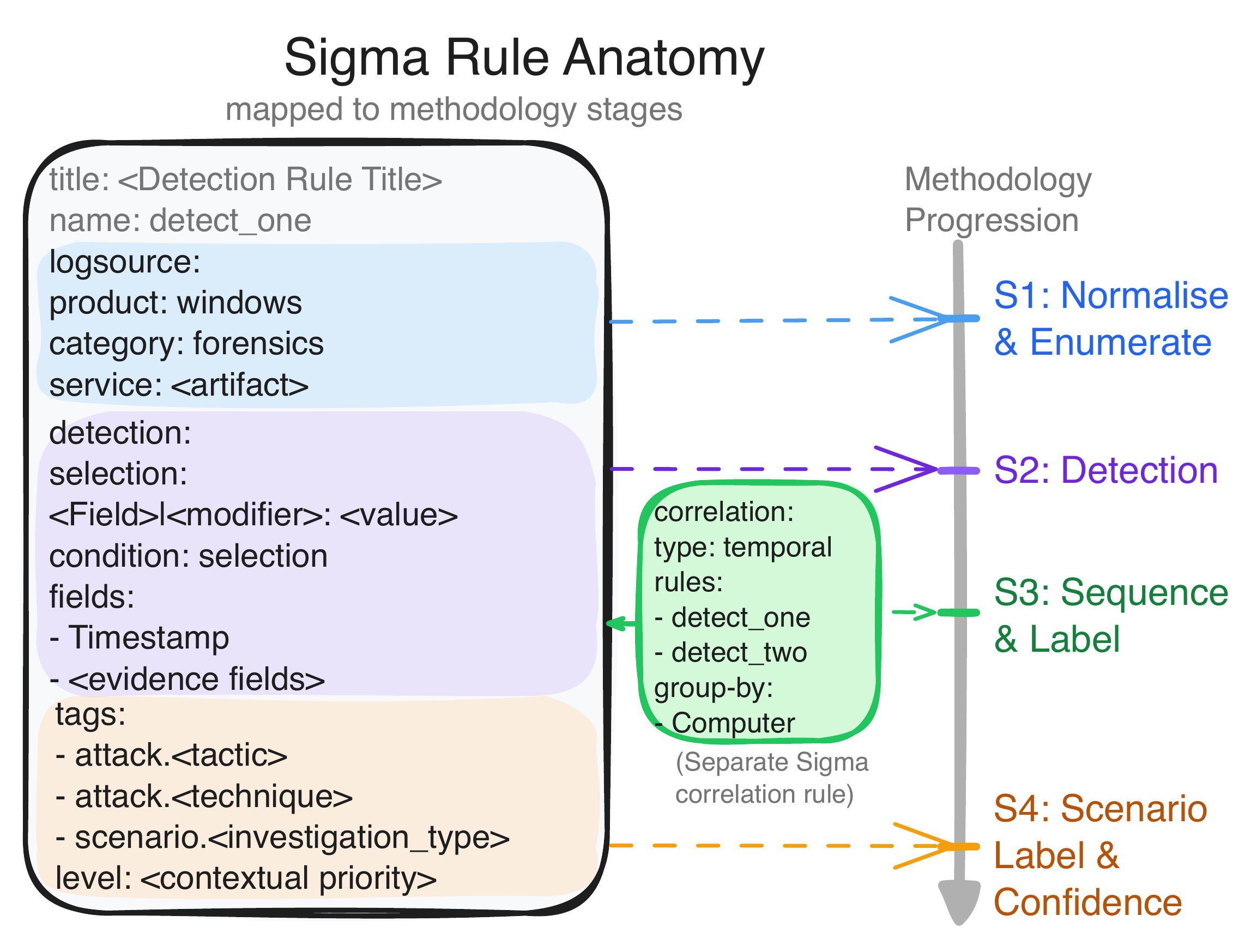}
    \caption{\textbf{Mapping methodology stages to the anatomy of a Sigma rule.}
    Each methodology stage maps to a specific part of the Sigma rule. Stage~1
    corresponds to the \texttt{logsource} section, which defines the forensic
    artefact. Stage~2 corresponds to the \texttt{detection} and \texttt{fields}
    sections. Stage~3 maps to a separate Sigma correlation rule that sequences
    base rule hits into an attack chain. Stage~4 maps to \texttt{tags} and
    \texttt{level}: \texttt{tags} labels the bundle with its ATT\&CK scenario
    and \texttt{level} records the bundle's priority for the case type under
    investigation.
    }
    \label{fig:diagram}
\end{figure}

The Velociraptor Unified Detection-Forensics Methodology is a four-stage
methodology that treats detection and forensics as one continuous
process running on the endpoint, as illustrated in
Figure~\ref{fig:diagram}. Its primary output is an evidence-backed
detection: an alert that is immediately supported by concrete forensic
artifacts such as Prefetch files, the NTFS USN journal, or WMI
repository state
\cite{velociraptor_2025_windowsforensicsprefetch,microsoft_using_2021}. Whether
a detection is implemented through Sigma rules
\cite{sigmahq_about_nd}, VQL, or custom logic, it is only considered
complete when it comes with its own proof.

The methodology is designed around four goals:
\begin{enumerate}
    \item Forensic-aware detection. Every rule is tied to artifacts
      that can be independently checked.
    \item High-confidence alerts. Correlation across multiple
      artifacts is used to reduce false positives.
    \item On-host automation. Velociraptor executes VQL and Sigma
      logic on the client to keep correlation close to the data.
    \item Repeatable investigations. The same artefact-backed rules
      can be reused across different threats and cases.
\end{enumerate}

Subsequent outputs, including reconstructed attack chains and
scenario labels, are important but secondary. They enrich and explain
detections that have already been forensically justified
\cite{luciano_digital_2018}.

The methodology assumes two input layers. The first is forensic data
sources, including execution-trace artifacts (for example, Prefetch
and program histories), configuration and persistence artifacts (such
as WMI objects), change-trace artifacts (such as USN journal records
and registry changes), and contextual data (such as host identity,
user, and time)
\cite{velociraptor_2025_windowsforensicsprefetch,velociraptor_2025_windowsforensicsusn}. The
second is detection logic, which consists of Sigma rules mapped to
Velociraptor log sources \cite{velociraptor_2025_sigma} and VQL
artifacts that implement filtering, joins, and correlations on-host.

It is assumed that endpoints run a Velociraptor client capable of
executing VQL locally and returning filtered results. This model also
assumes that, even if Windows Event Logs are missing or have been
cleared/disrupted, some separate OS-maintained artifacts (e.g.,
Prefetch/USN) may still be available for triage. Under these
conditions, the methodology can still produce evidence-backed
detections from durable artefacts.

More precisely, the methodology targets post-compromise triage on a
host that is already known or suspected to be compromised, rather than
real-time prevention or network-edge detection. The adversary is
assumed to hold administrative or SYSTEM-level privileges on that
host, since this is the privilege level required to clear the Windows
Event Log (T1070.001)~\cite{heiligenstein_2020_indicator} that the
methodology is designed to remain resilient against. In scope is
detection that survives event-log clearing or disruption. Out of
scope, and acknowledged as a limitation in
Section~\ref{sec:limitations}, is tampering with the durable artifacts
themselves (Prefetch, USN journal, and WMI repository), which the
methodology assumes are still present at triage time.

\subsection{Stage 1: Establish Factual Baseline}

Stage 1 builds a factual baseline of what exists on the host and what
has happened recently that is more resilient to Windows Event Log
clearing/disruption, independent of any suspicion. It is similar to
forensic triage, but automated and standardised through Velociraptor
\cite{casey_triage_2013}.

Using VQL, Velociraptor artifacts are run to enumerate persistent
state and extract historical traces of execution and change (for
example, Prefetch entries, USN records, or WMI repository objects)
\cite{velociraptor_2025_windowsforensicsprefetch}. Velociraptor
converts these artifacts into structured tables so they can be
analysed like logs.

The output is a set of baseline tables per host that describe recent
executions, file changes, and persistent configurations in a stable,
repeatable way. These tables give later stages a clean, log-like
structure for running detection rules.

\subsection{Stage 2: Correlate Specific Evidence}

Stage 2 is the core of the methodology. Its purpose is to transform
baseline findings from Stage 1 into evidence-backed detections by
correlating suspicious artifacts with independent proof of activity.

Conceptually, Stage 2 applies rule logic (Sigma or VQL conditions) to
the baseline tables to identify candidates of interest, such as an
unusual WMI subscription or a binary that appeared and ran from a
user-writable path \cite{sigmahq_about_nd}. For each hit, it triggers
targeted follow-up collection on the same host to confirm or refute
the suspicion, which may include Prefetch evidence showing that the
binary executed, process lists or historical process data confirming a
matching process, or other artifacts from the same time window.

Velociraptor enables this through on-host joins and chained artifacts
\cite{velociraptor_2025_sigma}. A Sigma rule mapped to a Velociraptor
log source is evaluated on the endpoint, and if it matches, additional
VQL queries can collect further context before the results are
returned.

The output of Stage 2 is a triage record or detection entry that
includes the rule that fired (a Sigma or VQL condition), the
suspicious artifact-at-rest such as a WMI object, Prefetch entry, or
USN record, and the corroborating artifacts that confirm execution or
impact. The Sigma/VQL rules and their associated BaseVQL log-source
mappings validated at this stage are reusable that can be deployed
directly in Velociraptor and can be tuned for future investigations.

At this point, the methodology has produced its primary deliverable: a
detection that already contains its own forensic justification.

\subsection{Stage 3: Analyse the Attack Chain}

Stage 3 takes evidence-backed detections from Stage 2 and organises
them into a minimal attack chain per host, describing attacker
progression rather than isolated events. Velociraptor gathers focused
timeline data, correlates events by time and context, and maps each
step to the corresponding MITRE ATT\&CK tactic and technique
\cite{mitre_attack_nd} to produce a standardised description of what
occurred at each phase, such as delivery, execution, persistence, or
defence evasion.

The output is a compact, reviewable attack chain with ATT\&CK labels
on each step. This chain expands individual Stage 2 detections into a
coherent sequence showing how the attacker progressed through the
system.

\subsection{Stage 4: Scenario Labelling and Confidence}
\label{sec:stage4}

Stage 4 does two things with the validated detections from Stage 3. First, it
groups them into a named rule bundle tied to MITRE ATT\&CK, for example a
``Payload Deployment'' bundle covering Ingress Tool Transfer (T1105), User
Execution (T1204.002), and Signed Binary Proxy Execution (T1218). Second, it
sets the Sigma \texttt{level} field on the bundle to record how strongly the
rule set applies to the case type under investigation.

The bundle name lets an analyst pick the right rule set by case type rather
than running every rule against every host. A ransomware response, for
example, typically prioritises rules for file encryption activity, shadow
copy deletion, and backup tampering. An intellectual property theft
investigation prioritises a different set: credential access, data staging,
and exfiltration over network or removable media. Stage 4 makes this filtering
explicit by tagging rules with the case types they are useful for, so the rule
library can be applied selectively instead of in full.

The \texttt{level} value is a qualitative judgement by the detection author
on how strongly the rule set applies to the case type under investigation.
It is set per bundle so that downstream analysts can filter the rule library
by investigative priority. The two case studies in
Sections~\ref{sec:resilient} and~\ref{sec:confidence} show how this value is
chosen in practice.
\section{Case Study 1: Artifact-Based Triage When Event Logs Are Cleared (Prefetch + USN)}
\label{sec:resilient}

\noindent\textbf{Scenario.} This case study evaluates a triage-phase workflow in
which VQL filtering \& Sigma rules are executed through Velociraptor's \texttt{Windows.\allowbreak Sigma.\allowbreak BaseVQL}
model over historical forensic artefacts rather than live event streams. In a
controlled VMware lab, a payload (\texttt{Edge.exe}) was generated with \texttt{msfvenom},
delivered via \texttt{certutil}, executed on a Windows 10 victim, and followed
by privilege escalation to \texttt{NT AUTHORITY\textbackslash SYSTEM}. The goal
is to show that, when Windows Event Logs are unavailable, misconfigured, or
cleared/disrupted (T1070.001), Prefetch and USN artefacts may still support
evidence-backed triage. Detections were implemented using the forensic
log sources i.e Prefetch \texttt{forensics/windows/prefetch} and
USN journal \texttt{forensics/windows/usn}.
\cite{cohen_2025_developing,heiligenstein_2020_indicator,microsoft_2025_certutil}.

\subsection{Stage 1: Establish Factual Baseline}

Stage 1 builds a timeline-first baseline from two OS-maintained artefacts:
Prefetch, which provides execution-of-record evidence, and the NTFS USN Change Journal,
which provides change-of-record evidence \cite{velociraptor_2025_windowsforensicsprefetch,velociraptor_2025_windowsforensicsusn,microsoft_using_2021}. Velociraptor normalises both into BaseVQL tables suitable for
Sigma evaluation \cite{velociraptor_2025_sigma}. In this case study,
Prefetch is used to identify executed binaries, paths, run counts,
and last-run timestamps, while USN is used to identify \texttt{.exe}/\texttt{.dll}
create or rename activity in user-writable locations such as \texttt{Downloads}
and \texttt{AppData\textbackslash Local\textbackslash Temp} \cite{velociraptor_2025_windowsforensicsprefetch,velociraptor_2025_windowsforensicsusn}. The resilience claim here is limited to Windows Event Log clearing/disruption:
these artefacts are independent of the event-log stream, although they are not tamper-proof

\subsection{Comparison: What Traditional Logs Show vs. Triage Artifacts (Balanced View)}

Log-centric investigations typically reconstruct this chain using
process creation and file create telemetry (e.g., Security 4688 or
Sysmon EIDs 1/3/11) \cite{russinovich_2024_sysmon}. When those logs
are present and correctly collected, they provide richer context such
as full command lines, parent/child relationships, and network
metadata. However, in triage, it is common that this telemetry was
never deployed, was misconfigured, or is unavailable due to
clearing/disruption (T1070.001). Under those conditions, event-log
queries may show little beyond secondary indicators (e.g. evidence
consistent with log interruption) and may miss the concrete
drop-and-execute timeline. In contrast, Prefetch and USN are
OS-maintained artifacts that are independent of the event-log stream
and may still provide execution-of-record and change-of-record
evidence after event log clearing/disruption. Clearing logs does not
delete them; removing them typically requires additional privileged
actions beyond log clearing.

A qualitative comparison with Hayabusa showed the same dependence on
EVTX log availability. When the relevant Windows Event Log files were
intact, Hayabusa was able to identify the user access enumeration
activity from the preserved EVTX records. However, after the EVTX logs
were deleted, Hayabusa no longer exposed the underlying enumeration
behaviour and instead only indicated that the event logs had been cleared.
This reinforces the distinction between log-centric and artefact-centric
visibility in this case study: log-based tools can provide strong
behavioural detection when EVTX data is present, but once those logs are removed,
visibility may collapse to the anti-forensic action itself rather
than the original attacker activity.

\subsection{Stage 2: Correlate Specific Evidence (Sigma over BaseVQL)}

\textbf{Objective.} Convert artifact collection to evidence-backed
detections by evaluating compact Sigma rules at the endpoint against
BaseVQL-backed forensic log sources
\cite{cohen_2025_developing,velociraptor_2025_sigma}.

\subsubsection{Sigma rules used (exact YAML)}
\textbf{Rule A --- Prefetch: LOLBins present (forensics)}
\begin{lstlisting}
LET RuleA ='''
title: Prefetch - LOLBins present (forensics)
status: experimental
logsource:
  category: forensics
  product: windows
  service: prefetch
detection:
  sel:
    Image|re: '(?i)^(powershell|edge|cmd|rundll32|certutil|msiexec)\.exe$'
  condition: sel
level: medium
'''
SELECT * FROM Artifact.Windows.Sigma.BaseVQL(SigmaRules=RuleA)
\end{lstlisting}

\textbf{Rule B --- USN: EXE/DLL in user Downloads/Temp }
\begin{lstlisting}
LET RuleB ='''
title: USN – EXE/DLL created or renamed in user Downloads/Temp
status: experimental
logsource:
  category: forensics
  product: windows
  service: usn
detection:
  staging_path:
    OSPath|re: '(?i)[/\\]Users[/\\][^/\\]+[/\\](Downloads|AppData[/\\]Local[/\\]Temp|Temp)([/\\]|$)'
  bin_ext:
    FileName|re: '(?i)\.(exe|dll)$'
  condition: staging_path and bin_ext
level: medium
'''
SELECT * FROM Artifact.Windows.Sigma.BaseVQL(SigmaRules=RuleB)
\end{lstlisting}
\subsubsection{Outcome (rule-level)}
Rule B (USN) provides staging evidence that an executable/DLL appeared in common
user-writable directories (e.g., \texttt{Downloads} and \texttt{Temp}) (T1105
context). Rule A (Prefetch) provides execution-of-record evidence for commonly
abused or high-signal binaries (e.g., \texttt{certutil.exe},
\texttt{rundll32.exe}, \texttt{msiexec.exe}, \texttt{powershell.exe}), including
last-run timestamps and run counts (T1204.002 context). When the observed
execution involves signed binaries commonly used for proxy execution, the chain
can also align with Signed Binary Proxy Execution (T1218)
\cite{homewood_2017_ingress,gaardls_2018_signed}. Stage~3 correlates USN staging
with Prefetch execution traces to reconstruct a compact drop$\rightarrow$execute
sequence suitable for triage.

\subsection{Stage 3: Analyse the Attack Chain}

Stage 3 reconstructs the minimal drop-to-execute chain by correlating USN
staging events with subsequent Prefetch execution evidence. The reconstruction
follows two steps: first, identify executable or DLL creation/rename activity
under \texttt{Downloads} or \texttt{Temp}; second, identify Prefetch evidence
for the same executable, or for proximate signed binaries used in proxy execution \cite{velociraptor_2025_windowsforensicsusn,velociraptor_2025_windowsforensicsprefetch,gaardls_2018_signed}. Table~\ref{tab:cs1-prefetch-hits}
shows the key execution-of-record hits used for reconstruction, including \texttt{CERTUTIL.EXE},
\texttt{EDGE.EXE}, \texttt{RUNDLL32.EXE}, and \texttt{MSIEXEC.EXE}. In the same time window,
USN also records a short burst of DLL staging in a GUID-named \texttt{Temp} directory,
consistent with transient post-execution staging. Taken together, the artefacts support
a compact sequence of delivery, payload execution, and proxy execution.

\subsubsection*{VQL used to extract Prefetch execution hits (triage scoping)}
\textbf{Syntax}
\begin{lstlisting}
SELECT Executable, RunCount, LastRunTimes, Binary
FROM Artifact.Windows.Forensics.Prefetch()
WHERE
  --- user-writable paths
  Binary =~ '''(?i)\\Users\\[^\\]+\\(Downloads|AppData\\Local\\Temp|Temp)\\'''
\end{lstlisting}
\subsubsection*{VQL used to extract USN staging hits}
\textbf{Syntax}
\begin{lstlisting}
SELECT Timestamp, FileName, OSPath, Reason, MFTId, ParentMFTId
FROM Artifact.Windows.Forensics.Usn()
WHERE OSPath =~ '(?i)\\Users\\.*\\(Downloads|AppData\\Local\\Temp|Temp)\\'
  AND FileName =~ '(?i)\.(exe|dll)$'
ORDER BY Timestamp DESC
LIMIT 300
\end{lstlisting}
\begin{table*}[t]
  \caption{CS1 Prefetch Execution Evidence (selected hits used for chain reconstruction)}
  \label{tab:cs1-prefetch-hits}
  \centering
  \scriptsize
  \begingroup
  \setlength{\tabcolsep}{5pt}
  \renewcommand{\arraystretch}{1.25}
  \begin{tabularx}{\textwidth}{@{}p{2.2cm}p{1.7cm}>{\raggedright\arraybackslash}p{5.2cm}>{\raggedright\arraybackslash}X@{}}
    \toprule
    \normalfont\textbf{Timestamp (UTC)} & \normalfont\textbf{Filename} &
    \normalfont\textbf{Path} & \normalfont\textbf{Details (Prefetch corroboration)} \\
    \midrule
    2025-09-24 04:12:33.598Z & CERTUTIL.EXE &
    \raggedright\path|\Windows\System32\certutil.exe| &
    pf=\texttt{CERTUTIL.EXE-57AECE36.pf}; \textbf{RunCount=2} (two executions within 04:11--04:12 window); LastRunTime=\texttt{2025-09-24T04:12:33.598Z}. \\[3pt]
    2025-09-24 04:14:42.833Z & EDGE.EXE &
    \raggedright\path|\Users\Aditya\Downloads\Edge.exe| &
	  pf=\texttt{EDGE.EXE-1E462A13.pf}; evidence of payload execution from a user-writable directory (Downloads); LastRunTime=\texttt{2025-09-24T04:14:42.833Z}. \\[3pt]
    2025-09-24 04:26:30.528Z & RUNDLL32.EXE &
    \raggedright\path|\Windows\System32\rundll32.exe| &
    pf=\texttt{RUNDLL32.EXE-EED2899C.pf}; proxy-execution context (signed LOLBin); LastRunTime=\texttt{2025-09-24T04:26:30.528Z}. \\[3pt]
    2025-09-24 04:27:10.968Z & MSIEXEC.EXE &
    \raggedright\path|\Windows\System32\msiexec.exe| &
    pf=\texttt{MSIEXEC.EXE-7D20CFB0.pf}; proxy-execution context (T1218.007); LastRunTime=\texttt{2025-09-24T04:27:10.968Z}. \\
    \bottomrule
  \end{tabularx}
  \endgroup
\end{table*}

\subsection{Stage 4: Scenario Labelling and Confidence}

Stage 4 maps the correlated artefacts to defensible ATT\&CK-labelled behaviour.
USN staging in user-writable directories is consistent with Ingress Tool
Transfer (T1105), Prefetch execution from those locations supports User
Execution (T1204.002), and Prefetch traces for signed binaries such as
\texttt{rundll32.exe}, \texttt{msiexec.exe}, and \texttt{certutil.exe} support
Signed Binary Proxy Execution (T1218)
\cite{homewood_2017_ingress,gaardls_2018_signed,mitre_attack_nd}.

\subsection{Data-Volume Cost: Sysmon vs.\ Artefact-Based Detection}

For the CS1 evaluation, Sysmon volume was measured from the
\texttt{Microsoft-Windows-Sysmon/Operational} log on the Windows test machine
over a 24-hour period. At the end of the measurement window, the log size was
2{,}166{,}784 bytes across 1{,}427 records, equivalent to approximately
2.17~MB/day per endpoint (2.07~MiB/day). Extrapolated linearly to
10{,}000 endpoints, this corresponds to approximately 21.67~GB/day of Sysmon
telemetry (20.18~GiB/day) that would need to be transported and stored
centrally. This baseline is conservative because Sysmon volume depends on host
activity and configuration \cite{russinovich_2024_sysmon}.

The raw artefact exports from CS1 illustrate why endpoint-side processing is
important. The raw USN JSON export was 10.10~MB across 22{,}781 rows, reflecting
the fact that the USN journal is a general OS-maintained record of file-system change on the volume.
By contrast, the full Prefetch JSON export was only 134.9~KB across 189 rows,
and already contained the key CS1 execution evidence such as
\texttt{CERTUTIL.EXE} and \texttt{EDGE.EXE}. Crucially, these raw export sizes
describe local artefact processing, not the network cost of the detection
workflow itself. In operational use, BaseVQL returns only matched Prefetch/USN
rows rather than the full underlying artefact stream, so the transmitted
results remain much smaller than continuous Sysmon forwarding.

Accordingly, the advantage of the artefact-based approach in CS1 is not that it
detects something Sysmon cannot, but that it shifts filtering to the endpoint
and transmits only evidence-backed hits. For triage-at-scale, this provides a
more data-efficient operating point than forwarding standing Sysmon telemetry
from every host, while still preserving useful forensic detection capability
for the drop-to-execute scenario.

\subsection{Takeaways}

Within this single-host scenario, the case study shows what becomes visible
when Sigma rules operate over durable forensic artefacts rather than only live logs. USN provides staging evidence, Prefetch
provides execution confirmation, and their correlation yields an evidence-backed
detection bundle that remains useful when Windows Event Logs are unavailable or cleared.
The result is a shorter, more reviewable reconstruction of attacker activity while
preserving the paper's four-stage methodology.

\section{Case Study 2: Artefact-Based WMI Persistence Monitoring and Correlation}
\label{sec:confidence}

Windows Management Instrumentation (WMI) event subscriptions provide a
mechanism for executing code in response to system events and are
commonly abused for persistence \cite{murphy_2020_event}. A permanent
subscription consists of three objects stored in the WMI repository:
an \texttt{\_\_EventFilter} defining the trigger condition, an
\texttt{EventConsumer} specifying the action, and a
\texttt{FilterToConsumerBinding} linking the two. Because these objects
are durable configuration artefacts rather than transient processes,
they persist across reboots and may execute with elevated privileges.

Detection is challenging because Windows does not reliably log the
creation of permanent WMI subscriptions. SIEM architectures relying on
centralised log forwarding may therefore miss persistence configuration
events if telemetry is unavailable at compromise time
\cite{graeber_2015_abusing}. This case study demonstrates an
artefact-based approach in which Velociraptor enumerates WMI repository
objects locally, applies Sigma-driven filtering, correlates execution
evidence, and produces structured investigative outputs. When scheduled
periodically, this model supports continuous monitoring while
maintaining low telemetry volume.

\subsection{Scenario, Testbed, and SIEM Comparison}

This experiment was conducted in a controlled VMware laboratory with a
Windows~11~x64 victim and a separate attacker host. Persistence was
established using Atomic Red Team test T1546.003-1
(\emph{Persistence via WMI Event Subscription --
CommandLineEventConsumer}), which creates a permanent
\texttt{\_\_EventFilter}, \texttt{CommandLineEventConsumer}, and
\texttt{FilterToConsumerBinding}. This reproduces adversary tradecraft
associated with MITRE ATT\&CK sub-technique T1546.003.

Unlike traditional SIEM architectures that forward large volumes of
endpoint telemetry to a central server, Velociraptor executes detection
logic locally using the
\texttt{Artifact.\allowbreak Windows.\allowbreak Sigma.\allowbreak BaseVQL} model. Only rule matches are
returned, significantly reducing network and storage overhead
(Figure~\ref{fig:velociraptor_sigma_flow}). Although demonstrated here
in a triage context, the same artefact queries can be scheduled
periodically to support continuous monitoring of WMI persistence state.

\subsection{Stage 1: Identifying WMI Persistence Artefacts}

Stage~1 establishes a baseline by enumerating permanent WMI subscription
objects. Each subscription contains three components:
\texttt{\_\_EventFilter}, \texttt{\_\_EventConsumer}, and
\texttt{FilterToConsumerBinding}. The filter defines the triggering WQL
query, the consumer specifies the payload, and the binding enables
execution when the condition is met
\cite{murphy_2020_event,marcho_2019_wmi}.

Rather than relying on event logs, Velociraptor directly queries the WMI
repository using built-in artefacts. This approach identifies persistence
even when logging telemetry is absent or incomplete
\cite{graeber_2015_abusing}. When executed periodically, the enumeration
provides continuous visibility into subscription state while returning
only structured artefact data.

\subsection{Stage 2: Applying Sigma-Based Artefact Filtering}

Stage~2 converts raw artefact enumeration into structured detection
logic. The WMI objects identified in Stage~1 are evaluated using Sigma
rules translated into Velociraptor VQL queries. This analysis extracts
the WQL query from \texttt{\_\_EventFilter}, examines execution logic
in \texttt{\_\_EventConsumer}, and confirms the active
\texttt{FilterToConsumerBinding}
\cite{velociraptor_overview_nd,ballenthin_2015_whymi}.
Detection logic executes locally on the endpoint, and only
rule matches are returned, reducing data volume compared
to centralised log forwarding.

In this case study, the adversary deployed an uptime-gated filter using
\texttt{\_\_Instance\allowbreak Modification\allowbreak Event} and the
\texttt{Win32\_\allowbreak PerfFormattedData\_\allowbreak PerfOS\_\allowbreak System} provider. The WQL
condition triggered when \texttt{SystemUpTime} exceeded a threshold,
delaying execution until shortly after system startup. This behaviour
aligns with MITRE ATT\&CK technique T1546.003
\cite{murphy_2020_event}.

Operational detection is achieved through a Sigma rule executed within
Velociraptor’s Curated BaseVQL framework. A custom WMI log source,
developed as part of this work, allows Sigma rules to operate directly
on WMI persistence artefacts rather than abstract log events.

\subsubsection{Sigma Rule Used in This Case Study}

\textbf{Sigma Rule --- WMI: Uptime-Gated \_\_InstanceModificationEvent Filter}

\begin{lstlisting}
title: WMI Uptime-Based __InstanceModificationEvent Filter

logsource:
  product: windows
  service: wmi

detection:
  q_event:
    EventData.FilterDetails.Query|contains: "__InstanceModificationEvent"
  q_uptime:
    EventData.FilterDetails.Query|contains: "SystemUpTime"
  condition: q_event and q_uptime

level: high
tags:
  - attack.persistence
  - attack.t1546.003
\end{lstlisting}

For brevity, non-essential Sigma metadata fields
(e.g., description, references, and dates) are omitted.
Only detection-relevant elements are shown.

Within Velociraptor, the rule is evaluated using the
\texttt{Artifact.\allowbreak Windows.\allowbreak Sigma.\allowbreak BaseVQL} artefact. For
continuous monitoring, the query is scheduled every
60 seconds using \texttt{clock()}:

\begin{lstlisting}
SELECT * FROM foreach(
  row={ SELECT * FROM clock(period=60) },
  query={
    SELECT *
    FROM Artifact.Windows.Sigma.BaseVQL(
      SigmaRules=RulesWMI
    )
  }
)
\end{lstlisting}

\subsubsection{Sigma--Forensics Integration via the
\texttt{*/windows/wmi} Log Source}

\begin{lstlisting}
*/windows/wmi = {
  SELECT
    timestamp(epoch=now()) AS Timestamp,
    dict(
      Computer = Hostname,
      Channel = 'Velociraptor',
      TimeCreated = dict(SystemTime = now())
    ) AS System,
    dict(
      Namespace = Namespace,
      FilterDetails = FilterDetails,
      ConsumerDetails = ConsumerDetails
    ) AS EventData
  FROM Artifact.Windows.Persistence.PermanentWMIEvents(
    AllRootNamespaces = false
  )
},
\end{lstlisting}

This log source exposes WMI persistence artefacts in a
Sigma-compatible schema. When both
\texttt{\_\_Instance\allowbreak Modification\allowbreak Event} and
\texttt{SystemUpTime} appear in a filter query, the detection
triggers forensic enrichment such as Prefetch analysis and
process lineage correlation.

\subsection{Stage 3: Execution Correlation and Attack Chain Reconstruction}

Stage~3 validates that the persistence configuration has
resulted in execution. When the WQL condition is satisfied,
the WMI Provider Service (\texttt{WmiPrvSe.exe}) launches
the configured payload, typically under the SYSTEM account
\cite{murphy_2020_event}.

Velociraptor correlates WMI subscription artefacts with
process lineage and Prefetch execution records. In this
case study, Prefetch analysis confirmed that
\texttt{notepad.exe} executed approximately four minutes
after system startup, matching the uptime-gated WQL
condition. Process lineage identified \texttt{WmiPrvSe.exe}
as the parent process, providing high-confidence attribution
of WMI-triggered execution.

This correlation produces a clear attack chain:
(1) WMI subscription creation,
(2) uptime-gated trigger condition,
(3) execution via \texttt{WmiPrvSe.exe},
(4) SYSTEM-level payload launch.
The detection logic is reusable across WMI persistence
variants by modifying filter or consumer conditions.

\subsection{Stage 4: Scenario Labelling and Confidence}

Stage~4 converts detection output into structured
investigative categories. Triggered rules are classified
according to investigative scenarios.

\textbf{Input.}
Correlated results from Stage~3 including WMI artefacts,
execution via \texttt{WmiPrvSe.exe}, and supporting
Prefetch evidence.

\textbf{Processing.}
The rule is categorised within a
\emph{Persistence Establishment Scenario} aligned with
MITRE ATT\&CK T1546.003.

\textbf{Output.}
A labelled rule describing the scenario:
``WMI Persistence – Uptime-Gated Subscription
(High Priority for Persistence Investigations).’’

This structured categorisation supports prioritisation,
cross-case comparison, and repeatable investigative
workflows.

\subsection{Discussion: Detection Coverage and Telemetry Volume}

This case study compares Sysmon-based logging with
Velociraptor’s artefact-driven monitoring for WMI persistence.

\textbf{Sysmon visibility.}
Sysmon can detect WMI persistence via Event ID~19, 20, and 21,
but these are not enabled by default and require explicit
configuration.
In testing, Atomic Red Team T1546.003-1 produced no WMI-related
events under default settings, demonstrating that persistence
may remain undetected without prior instrumentation.

\textbf{Telemetry volume.}
The Sysmon log reached approximately 108~MB during testing,
equating to roughly 2.5--3~GB per day per endpoint. Most data
consisted of high-frequency process and network events, with
WMI-related signals forming only a small fraction. This creates
a high-noise environment where large volumes of benign telemetry
must be processed to identify rare persistence activity.

\textbf{Artefact-based monitoring.}
Velociraptor instead queries WMI repository artefacts directly
and applies Sigma rules locally. Using periodic execution via
\texttt{clock(period=60)}, only matching results are returned,
eliminating continuous log forwarding and significantly reducing
data volume.

\textbf{Summary.}
Sysmon detection depends on configuration and timing, whereas
Velociraptor provides complete coverage by analysing persistent
artefacts. This reduces telemetry from gigabytes per day to a
small number of evidence-backed detections, illustrating the
efficiency of endpoint-side processing in Case Study~2.

\subsection{Takeaways}

Within this single-host WMI persistence scenario, this case study
illustrates how the four-stage methodology produces evidence-backed
detections by correlating WMI repository artefacts with Prefetch
execution evidence and SYSTEM-level process lineage. The \texttt{/windows/wmi}
log source enables Sigma rules to operate directly on
forensic-quality WMI persistence data, creating a
low-volume detection workflow that prioritises the rule set
for persistence investigations.




\section{Discussion}
\label{sec:discussion}

The four-stage methodology illustrates how detection engineering and
forensic acquisition can operate as a single workflow. Both
case studies show that Sigma rules operating over durable
artifacts-Prefetch, USN, and WMI repository-can produce
evidence-backed detections that remain effective after Windows Event
Log clearing/disruption (T1070.001). If Prefetch/USN evidence is also
missing or appears reset in a way inconsistent with host activity,
that condition can itself be treated as evidence consistent with
additional anti-forensic behaviour beyond event-log clearing.

\subsection{Evidence-Backed Detections and Log-Centric Workflows}

Both case studies illustrate the shift from alerts that rely on
volatile logs to detections that already include their own supporting
evidence. In the resilient baseline case study, detections still fire
even if event logs are assumed cleared because forensic-backed
detection rules operate over durable execution and change-trace
artifacts such as Prefetch and the NTFS USN journal
\cite{velociraptor_2025_windowsforensicsprefetch,microsoft_using_2021}. The
resulting hit already summarises which binary ran, from where, and
when. In the WMI case study, the logic moves from noting that a
suspicious subscription exists to showing that the subscription
executed a specific command under a particular token shortly after
boot, based on the correlation of repository state with Prefetch and
pslist \cite{tilbury_2023_finding,murphy_2020_event}. Across both
scenarios, the Stage 2 triage record acts as a small evidence bundle
rather than a vague or incomplete alert.

\subsection{Unifying Detection Engineering and Forensic Practice in Velociraptor}

The four-stage methodology also shows that detection engineering and forensics
can operate from the same artifacts and logic. Sigma and VQL rules determine
when to alert and also guide what should be collected and how different sources
are connected \cite{sigmahq_about_nd}. In the baseline case study, Sigma rules
run against Prefetch and USN log sources in the Windows Base VQL model, allowing
detection logic that would usually require a SIEM to operate on forensic
artefacts directly on the host. In the WMI case study, the \texttt{/windows/wmi}
log source exposes Filters, Consumers and Bindings in a Sigma-compatible format,
and the resulting correlation with execution artefacts drives the full
detection-to-collection workflow. In both cases, the Sigma rules and log sources
are immediately reusable outputs that practitioners can deploy to other
Velociraptor instances.

In effect, Velociraptor works as both the detector and the forensic
collector. It performs correlation on the endpoint and returns only
evidence-rich findings, which reduces noise but keeps the detail
analysts need. This changes the workflow from checking a SIEM alert
and then using another DFIR tool to receiving an alert that already
contains the artifacts that would have collected.

\subsection{From Triage to Continuous Monitoring}
\label{subsec:triage-to-monitoring}

Once a Sigma rule has been produced through the four-stage methodology,
converting it to a continuous monitor requires only a thin wrapper around the
existing query. This contrasts with traditional SIEM rule authoring, where
analysts work only in a monitoring context and depend on whichever event logs
are forwarded to them, with no forensic visibility into durable
artefacts~\cite{nist_sp800_92}. Rules developed through the four-stage
methodology, however, start from artefact-level triage. This means the analyst
has access to richer evidence than is normally available, and those same rules
can then be reused for monitoring with minimal modification.

Figure~\ref{lst:monitoring-template} shows the conversion pattern. Three VQL
mechanisms handle the transition. The \texttt{clock()} event plugin drives
periodic re-execution at a configurable interval~\cite{velociraptor_vql_clock}.
The \texttt{foreach()} plugin wraps the one-shot triage query so that it re-runs
on each tick~\cite{velociraptor_vql_foreach}. The \texttt{dedup()} plugin
suppresses rows whose key has not changed since the previous cycle, so only new
or modified artefacts produce output~\cite{velociraptor_vql_dedup}. The Sigma
rule and its BaseVQL log source remain identical to the triage version. This
pattern applies to Windows, macOS, and Linux BaseVQL models by substituting the
appropriate platform artefact.

\begin{figure}[pos=tbp]
\begin{lstlisting}
LET Rules <= '''
<sigma_rule_yaml>
'''

SELECT * FROM dedup(
  key="DedupKey",
  query={
    SELECT * FROM foreach(
      row={
        SELECT * FROM clock(period=60)
      },
      query={
        SELECT *,
          <dedup_expression> AS DedupKey
        FROM Artifact.<Platform>.Sigma.BaseVQL(
          SigmaRules=Rules
        )
      }
    )
  }
)
\end{lstlisting}
\caption{Monitoring conversion template. The triage query
  is wrapped in periodic execution via
  \texttt{clock()} and deduplicated so only new or
  changed artefacts are forwarded.}
\label{lst:monitoring-template}
\end{figure}

Only detection matches are returned, not raw artefact contents or full telemetry
streams. This keeps data volume low compared with conventional log forwarding.

\subsection{Limitations and Trade-offs}
\label{sec:limitations}

The methodology presented here is designed to complement, not replace,
traditional log-based detection. Event logs capture authentication
context, user session information, and security events such as failed
logons, privilege escalation, and account lockouts that Prefetch, USN
Journal, and WMI artifacts cannot record \cite{nist_sp800_86}. The
optimal production configuration combines both approaches: artifacts
provide resilience specifically when Windows Event Logs are
cleared/disrupted or unavailable, while logs supply user attribution
and policy-level events that file system artifacts inherently
lack. Each trade-off reflects design choices optimised for
post-incident triage rather than deficiencies in the approach itself.

Detection latency varies by operational mode. Post-incident triage of
existing artifacts (Case Study 1) is inherently retrospective. While
event-based monitoring capabilities provide real-time alerting (Case
Study 2), artifact analysis introduces detection
latency. Additionally, post-hoc analysis inherently loses volatile
data such as memory contents, active network connections, and running
process states that would be available to real-time detection
approaches \cite{nist_sp800_86}.

Each artifact source imposes its own retention and coverage
constraints. Windows stores a maximum of 1,024 Prefetch files on
Windows 8 and later systems, with high system activity causing older
execution evidence to be purged in first-in-first-out order
\cite{microsoft_ir_guidebook,sans_isc_prefetch_2022}. Prefetch files
record only the last eight execution timestamps and capture only the
first ten seconds of application execution, meaning activity after
this window goes unrecorded \cite{microsoft_ir_guidebook}. The USN
Journal operates as a circular buffer with a 32 MB default size limit;
once this limit is reached, older entries are overwritten by new ones,
potentially causing loss of historical evidence
\cite{druva_data_anomalies,troopers_antiforensics_2025}. WMI
persistence can be created in any namespace, not just
\texttt{root\textbackslash subscription}, so detection approaches
checking only the default namespace will miss persistence mechanisms
elsewhere \cite{specter_subverting_sysmon_2022}. Beyond retention and
coverage, an adversary holding administrative or SYSTEM-level
privileges can also remove or tamper with the durable artefacts
themselves, which constitutes additional anti-forensic behaviour
beyond event-log clearing and is out of scope for this paper rather
than a fundamental limitation of the methodology.

\section{Conclusion}
\label{sec:conclusion}

This work presents three contributions to the integration of detection
engineering and digital forensics. First, it introduces a four-stage methodology
that guides the systematic development of forensic detections, converting
artefact knowledge into reusable and testable detection rules suitable for both
post-incident triage and live monitoring. Second, through two case studies using
Velociraptor BaseVQL log sources for Prefetch, USN, and WMI artefacts, it
illustrates how artefact-based detections can support targeted forensic
triage in the tested scenarios. This helps investigators identify and
prioritise compromised endpoints without full disk acquisition. Third, it provides supporting evidence that
periodic artefact analysis offers continuous live monitoring while substantially
reducing data volume compared to conventional endpoint logging.

A key advantage of this approach is resilience to log tampering. Because
forensic artefacts such as Prefetch files, USN journal entries, and WMI
repository objects exist outside the Windows Event Log stream, they may remain
available even after log clearing. Removing these artefacts typically requires
additional privileged actions beyond log deletion, making the resulting
detections more durable than those built on event logs alone. The case studies
confirm that correlating these artefacts enables investigators to determine what
executed, how it was triggered, and at what privilege level.

Stage 4 of the methodology groups validated detections into named
ATT\&CK-labelled rule bundles and records each bundle's priority for the
case type under investigation via the Sigma \texttt{level} field. This lets
analysts select the most relevant rule sets for a given investigation type,
such as ransomware response or intellectual property theft, rather than
applying all available rules indiscriminately.

Overall, this unified detection-forensics methodology supports more
resilient and more informed incident response. By making durable forensic
artefacts part of the initial detection process and linking technical findings
to investigative context, this work offers a foundation for future research
and operational improvement. With continued testing and rule development,
the approach can extend to additional artefact types and attack techniques
across a broader range of threat scenarios, helping organisations achieve
full forensic readiness.

\FloatBarrier

\section*{Declaration of generative AI and AI-assisted technologies in the manuscript preparation process}
During the preparation of this work the authors used ChatGPT GPT 5.3
(OpenAI) and Claude Sonnet 4.6 and Claude Opus 4.6 (Anthropic) in
order to refine prose for language clarity, edit and format draft
text, and assist with LaTeX typesetting. After using these tools, the
authors reviewed and edited the content as needed and take full
responsibility for the content of the published article.

\bibliography{references}

\begin{thebibliography}{46}
\expandafter\ifx\csname natexlab\endcsname\relax\def\natexlab#1{#1}\fi
\providecommand{\url}[1]{\texttt{#1}}
\providecommand{\href}[2]{#2}
\providecommand{\path}[1]{#1}
\providecommand{\DOIprefix}{doi:}
\providecommand{\ArXivprefix}{arXiv:}
\providecommand{\URLprefix}{URL: }
\providecommand{\Pubmedprefix}{pmid:}
\providecommand{\doi}[1]{\href{http://dx.doi.org/#1}{\path{#1}}}
\providecommand{\Pubmed}[1]{\href{pmid:#1}{\path{#1}}}
\providecommand{\bibinfo}[2]{#2}
\ifx\xfnm\relax \def\xfnm[#1]{\unskip,\space#1}\fi
\bibitem[{Bace(2000)}]{bace_intrusion_2000}
\bibinfo{author}{Bace, R.G.}, \bibinfo{year}{2000}.
\newblock \bibinfo{title}{Intrusion Detection}.
\newblock \bibinfo{publisher}{Macmillan Technical Publishing},
  \bibinfo{address}{United Kingdom}.
\bibitem[{Ballenthin et~al.(2015)Ballenthin, Graeber and
  Teodorescu}]{ballenthin_2015_whymi}
\bibinfo{author}{Ballenthin, W.}, \bibinfo{author}{Graeber, M.},
  \bibinfo{author}{Teodorescu, C.}, \bibinfo{year}{2015}.
\newblock \bibinfo{title}{Whymi so sexy? wmi attacks, real-time defense, and
  advanced forensic analysis}.
\newblock \URLprefix \url{https://media.defcon.org/}.
\bibitem[{Ban et~al.(2023)Ban, Takahashi, Ndichu and Inoue}]{ban_breaking_2023}
\bibinfo{author}{Ban, T.}, \bibinfo{author}{Takahashi, T.},
  \bibinfo{author}{Ndichu, S.}, \bibinfo{author}{Inoue, D.},
  \bibinfo{year}{2023}.
\newblock \bibinfo{title}{Breaking alert fatigue: {AI}-assisted {SIEM}
  framework for effective incident response}.
\newblock \bibinfo{journal}{Applied Sciences} \bibinfo{volume}{13},
  \bibinfo{pages}{6610}.
\newblock \URLprefix \url{https://doi.org/10.3390/app13116610}.
\bibitem[{Berger(2025)}]{troopers_antiforensics_2025}
\bibinfo{author}{Berger, S.}, \bibinfo{year}{2025}.
\newblock \bibinfo{title}{Anti-forensics}, in: \bibinfo{booktitle}{Troopers
  Conference 2025}.
\newblock \URLprefix
  \url{https://troopers.de/downloads/troopers25/TR25_Anti-Forensics_8KYNCU.pdf}.
  \bibinfo{note}{conference presentation}.
\bibitem[{Breitinger et~al.(2024)Breitinger, Hilgert, Hargreaves, Sheppard,
  Overdorf and Scanlon}]{breitinger_dfrws_2024}
\bibinfo{author}{Breitinger, F.}, \bibinfo{author}{Hilgert, J.N.},
  \bibinfo{author}{Hargreaves, C.}, \bibinfo{author}{Sheppard, J.},
  \bibinfo{author}{Overdorf, R.}, \bibinfo{author}{Scanlon, M.},
  \bibinfo{year}{2024}.
\newblock \bibinfo{title}{{DFRWS} {EU} 10-year review and future directions in
  digital forensic research}.
\newblock \bibinfo{journal}{Forensic Science International: Digital
  Investigation} \bibinfo{volume}{48}, \bibinfo{pages}{301685}.
\newblock \URLprefix \url{https://doi.org/10.1016/j.fsidi.2023.301685}.
\bibitem[{Casey(2013)}]{casey_triage_2013}
\bibinfo{author}{Casey, E.}, \bibinfo{year}{2013}.
\newblock \bibinfo{title}{Triage in digital forensics}.
\newblock \bibinfo{journal}{Digital Investigation} \bibinfo{volume}{10},
  \bibinfo{pages}{85--86}.
\newblock \URLprefix \url{https://doi.org/10.1016/j.diin.2013.08.001}.
\bibitem[{Cohen(2025)}]{cohen_2025_developing}
\bibinfo{author}{Cohen, M.}, \bibinfo{year}{2025}.
\newblock \bibinfo{title}{Developing sigma rules in velociraptor ::
  Velociraptor - digging deeper!}
\newblock \URLprefix
  \url{https://docs.velociraptor.app/blog/2025/2025-02-02-sigma/}.
\bibitem[{Di~Giorgio(2021)}]{digiorgio_detection_2021}
\bibinfo{author}{Di~Giorgio, P.}, \bibinfo{year}{2021}.
\newblock \bibinfo{title}{Detection engineering: Defending networks with
  purpose}.
\newblock \bibinfo{type}{Technical Report}. SANS Institute.
\newblock \URLprefix \url{https://sansorg.egnyte.com/dl/nElpZhPfHa}.
  \bibinfo{note}{gIAC GCIA Gold Certification paper, Advisor: J. Hally}.
\bibitem[{{Druva}(2024)}]{druva_data_anomalies}
\bibinfo{author}{{Druva}}, \bibinfo{year}{2024}.
\newblock \bibinfo{title}{Data anomalies settings}.
\newblock \bibinfo{howpublished}{Online documentation}.
\newblock \URLprefix
  \url{https://help.druva.com/en/articles/8513206-data-anomalies-settings}.
\bibitem[{Gaardløs et~al.(2018)Gaardløs, Maharjan, ~ and
  Hurd}]{gaardls_2018_signed}
\bibinfo{author}{Gaardløs, H.C.}, \bibinfo{author}{Maharjan, N.},
  \bibinfo{author}{~, P.}, \bibinfo{author}{Hurd, W.}, \bibinfo{year}{2018}.
\newblock \bibinfo{title}{Signed binary proxy execution, technique t1218 -
  enterprise | mitre att\&ck®}.
\newblock \URLprefix \url{https://attack.mitre.org/techniques/T1218/}.
\bibitem[{Graeber(2015)}]{graeber_2015_abusing}
\bibinfo{author}{Graeber, M.}, \bibinfo{year}{2015}.
\newblock \bibinfo{title}{Abusing windows management instrumentation (wmi) to
  build a persistent, asyncronous, and fileless backdoor matt graeber black hat
  2015}.
\newblock \URLprefix \url{https://blackhat.com/}.
\bibitem[{Graeber and Christensen(2022)}]{specter_subverting_sysmon_2022}
\bibinfo{author}{Graeber, M.}, \bibinfo{author}{Christensen, L.},
  \bibinfo{year}{2022}.
\newblock \bibinfo{title}{Subverting Sysmon: Application of a Formalized
  Security Product Evasion Methodology}.
\newblock \bibinfo{type}{Technical whitepaper}. SpecterOps.
\newblock \URLprefix
  \url{https://specterops.io/wp-content/uploads/sites/3/2022/06/Subverting_Sysmon.pdf}.
\bibitem[{Hargreaves et~al.(2024)Hargreaves, Nelson and
  Casey}]{hargreaves_abstract_2024}
\bibinfo{author}{Hargreaves, C.}, \bibinfo{author}{Nelson, A.},
  \bibinfo{author}{Casey, E.}, \bibinfo{year}{2024}.
\newblock \bibinfo{title}{An abstract model for digital forensic analysis tools
  - a foundation for systematic error mitigation analysis}.
\newblock \bibinfo{journal}{Forensic Science International: Digital
  Investigation} \bibinfo{volume}{48}, \bibinfo{pages}{301679}.
\newblock \URLprefix \url{https://doi.org/10.1016/j.fsidi.2023.301679}.
\bibitem[{Heiligenstein(2020)}]{heiligenstein_2020_indicator}
\bibinfo{author}{Heiligenstein, L.}, \bibinfo{year}{2020}.
\newblock \bibinfo{title}{Indicator removal on host: Clear windows event logs,
  sub-technique t1070.001 - enterprise | mitre att\&ck®}.
\newblock \URLprefix \url{https://attack.mitre.org/techniques/T1070/001/}.
\bibitem[{Homewood et~al.(2017)Homewood, Hedges, Wise, Page, ApparitionSec and
  Wee}]{homewood_2017_ingress}
\bibinfo{author}{Homewood, A.}, \bibinfo{author}{Hedges, J.},
  \bibinfo{author}{Wise, J.}, \bibinfo{author}{Page, J.},
  \bibinfo{author}{ApparitionSec}, \bibinfo{author}{Wee, M.},
  \bibinfo{year}{2017}.
\newblock \bibinfo{title}{Ingress tool transfer, technique t1105 - enterprise |
  mitre att\&ck®}.
\newblock \URLprefix \url{https://attack.mitre.org/techniques/T1105/}.
\bibitem[{Kent et~al.(2006)Kent, Chevalier, Grance and Dang}]{nist_sp800_86}
\bibinfo{author}{Kent, K.}, \bibinfo{author}{Chevalier, S.},
  \bibinfo{author}{Grance, T.}, \bibinfo{author}{Dang, H.},
  \bibinfo{year}{2006}.
\newblock \bibinfo{title}{Guide to Integrating Forensic Techniques into
  Incident Response}.
\newblock \bibinfo{type}{Technical Report} \bibinfo{number}{NIST Special
  Publication 800-86}. National Institute of Standards and Technology.
  \bibinfo{address}{Gaithersburg, MD}.
\newblock \URLprefix \url{https://csrc.nist.gov/pubs/sp/800/86/final},
  \DOIprefix\doi{10.6028/NIST.SP.800-86}.
\bibitem[{Kent and Souppaya(2006)}]{nist_sp800_92}
\bibinfo{author}{Kent, K.}, \bibinfo{author}{Souppaya, M.},
  \bibinfo{year}{2006}.
\newblock \bibinfo{title}{Guide to Computer Security Log Management}.
\newblock \bibinfo{type}{Technical Report} \bibinfo{number}{NIST Special
  Publication 800-92}. National Institute of Standards and Technology.
  \bibinfo{address}{Gaithersburg, MD}.
\newblock \URLprefix \url{https://csrc.nist.gov/pubs/sp/800/92/final},
  \DOIprefix\doi{10.6028/NIST.SP.800-92}.
\bibitem[{Lange(2023)}]{lange_detection_2023}
\bibinfo{author}{Lange, K.}, \bibinfo{year}{2023}.
\newblock \bibinfo{title}{Detection engineering explained}.
\newblock \URLprefix
  \url{https://www.splunk.com/en_us/blog/learn/detection-engineering.html}.
\bibitem[{Luciano et~al.(2018)Luciano, Baggili, Topor, Casey and
  Breitinger}]{luciano_digital_2018}
\bibinfo{author}{Luciano, L.}, \bibinfo{author}{Baggili, I.},
  \bibinfo{author}{Topor, M.}, \bibinfo{author}{Casey, P.},
  \bibinfo{author}{Breitinger, F.}, \bibinfo{year}{2018}.
\newblock \bibinfo{title}{Digital forensics in the next five years}, in:
  \bibinfo{booktitle}{Proceedings of the International Conference on
  Availability, Reliability and Security (ARES 2018)},
  \bibinfo{address}{Hamburg, Germany}. pp. \bibinfo{pages}{1--14}.
\newblock \URLprefix \url{https://doi.org/10.1145/3230833.3232813}.
\bibitem[{Marcho(2019)}]{marcho_2019_wmi}
\bibinfo{author}{Marcho, C.}, \bibinfo{year}{2019}.
\newblock \bibinfo{title}{Wmi: Rebuilding the wmi repository}.
\newblock \URLprefix
  \url{https://techcommunity.microsoft.com/blog/askperf/wmi-rebuilding-the-wmi-repository/373846}.
\bibitem[{Microsoft(2025)}]{microsoft_2025_certutil}
\bibinfo{author}{Microsoft}, \bibinfo{year}{2025}.
\newblock \bibinfo{title}{certutil}.
\newblock \URLprefix
  \url{https://learn.microsoft.com/en-us/windows-server/administration/windows-commands/certutil}.
\bibitem[{{Microsoft Learn}(2021)}]{microsoft_using_2021}
\bibinfo{author}{{Microsoft Learn}}, \bibinfo{year}{2021}.
\newblock \bibinfo{title}{Using the change journal identifier}.
\newblock \bibinfo{howpublished}{Online documentation}.
\newblock \URLprefix
  \url{https://learn.microsoft.com/en-us/windows/win32/fileio/using-the-change-journal-identifier}.
  \bibinfo{note}{accessed: 17 November 2025}.
\bibitem[{{Microsoft Security Response Center}(2025)}]{microsoft_ir_guidebook}
\bibinfo{author}{{Microsoft Security Response Center}}, \bibinfo{year}{2025}.
\newblock \bibinfo{title}{Incident response reference guide}.
\newblock \URLprefix
  \url{https://cdn-dynmedia-1.microsoft.com/is/content/microsoftcorp/microsoft/final/en-us/microsoft-brand/documents/IR-Guidebook-Final.pdf}.
\bibitem[{{{MITRE ATT\&CK}}(2025)}]{mitre_attack_nd}
\bibinfo{author}{{{MITRE ATT\&CK}}}, \bibinfo{year}{2025}.
\newblock \bibinfo{title}{{MITRE ATT\&CK}\textregistered}.
\newblock \bibinfo{howpublished}{Knowledge base}.
\newblock \URLprefix \url{https://attack.mitre.org/}.
\bibitem[{Murphy et~al.(2020)Murphy, French and Chaudhari}]{murphy_2020_event}
\bibinfo{author}{Murphy, B.}, \bibinfo{author}{French, D.},
  \bibinfo{author}{Chaudhari, V.}, \bibinfo{year}{2020}.
\newblock \bibinfo{title}{Event triggered execution: Windows management
  instrumentation event subscription, sub-technique t1546.003 - enterprise |
  mitre att\&ck®}.
\newblock \URLprefix \url{https://attack.mitre.org/techniques/T1546/003}.
\bibitem[{Myllyl{\"a}(2021)}]{myllyla_detecting_2021}
\bibinfo{author}{Myllyl{\"a}, J.}, \bibinfo{year}{2021}.
\newblock \bibinfo{title}{Detecting Cyber Attacks in Time: Combining Attack
  Simulation with Detection Logic}.
\newblock Master's thesis. University of Jyv{\"a}skyl{\"a}.
  \bibinfo{address}{Jyv{\"a}skyl{\"a}, Finland}.
\newblock \URLprefix
  \url{https://jyx.jyu.fi/bitstreams/b1088dd9-77be-4ade-8537-e739455326f5/download}.
\bibitem[{{Palo Alto Networks}(2026)}]{paloalto_dfir_nd}
\bibinfo{author}{{Palo Alto Networks}}, \bibinfo{year}{2026}.
\newblock \bibinfo{title}{What is digital forensics and incident response
  ({DFIR})?}
\newblock \URLprefix
  \url{https://www.paloaltonetworks.com.au/cyberpedia/digital-forensics-and-incident-response}.
\bibitem[{Ras(2018)}]{ras_digital_2018}
\bibinfo{author}{Ras, D.J.}, \bibinfo{year}{2018}.
\newblock \bibinfo{title}{Digital Forensic Readiness Architecture for Cloud
  Computing Systems}.
\newblock Master's thesis. University of Pretoria. \bibinfo{address}{Pretoria,
  South Africa}.
\bibitem[{Russinovich and Garnier(2024)}]{russinovich_2024_sysmon}
\bibinfo{author}{Russinovich, M.}, \bibinfo{author}{Garnier, T.},
  \bibinfo{year}{2024}.
\newblock \bibinfo{title}{Sysmon - windows sysinternals}.
\newblock \URLprefix
  \url{https://learn.microsoft.com/en-us/sysinternals/downloads/sysmon}.
\bibitem[{{SANS Institute}(2018)}]{sans_nextgen_siem}
\bibinfo{author}{{SANS Institute}}, \bibinfo{year}{2018}.
\newblock \bibinfo{title}{Evaluator's guide to next-gen siem}.
\newblock \URLprefix
  \url{https://www.sans.org/media/vendor/evaluator-039-s-guide-nextgen-siem-38720.pdf}.
\bibitem[{Shaaban and Sapronov(2016)}]{shaaban_practical_2016}
\bibinfo{author}{Shaaban, A.}, \bibinfo{author}{Sapronov, K.},
  \bibinfo{year}{2016}.
\newblock \bibinfo{title}{Practical Windows Forensics: Leverage the Power of
  Digital Forensics for Windows Systems}.
\newblock \bibinfo{publisher}{Packt Publishing}, \bibinfo{address}{Birmingham}.
\bibitem[{Shelke and Frantti(2025)}]{shelnt_exploring_2025}
\bibinfo{author}{Shelke, P.}, \bibinfo{author}{Frantti, T.},
  \bibinfo{year}{2025}.
\newblock \bibinfo{title}{Exploring the possibilities of {Splunk Enterprise
  Security} in advanced cyber threat detection}, in:
  \bibinfo{booktitle}{Proceedings of the 20th International Conference on Cyber
  Warfare and Security (ICCWS 2025)}, pp. \bibinfo{pages}{605--613}.
\newblock \URLprefix \url{https://doi.org/10.34190/iccws.20.1.3326}.
\bibitem[{Shetty(2022)}]{shetty_dark_2022}
\bibinfo{author}{Shetty, M.P.}, \bibinfo{year}{2022}.
\newblock \bibinfo{title}{Dark \& Deep Web: Advanced Forensic Analysis of {Tor}
  Browser and Implications for Law Enforcement Agencies}.
\newblock \bibinfo{type}{Technical Report}. Centre for Cybercrime Investigation
  Training \& Research, Data Security Council of India.
\newblock \URLprefix
  \url{https://www.dsci.in/files/content/knowledge-centre/2023/Advanced-Forensic-Analysis-of-Tor-Browser-and-Implications-for-Law-Enforcement-Agencies.pdf}.
\bibitem[{{SigmaHQ}(2025)}]{sigmahq_about_nd}
\bibinfo{author}{{SigmaHQ}}, \bibinfo{year}{2025}.
\newblock \bibinfo{title}{About {Sigma} {\textbar} {Sigma} detection format}.
\newblock \bibinfo{howpublished}{Project documentation}.
\newblock \URLprefix \url{https://sigmahq.io/docs/guide/about.html}.
\bibitem[{Soltani and Hosseini~Seno(2023)}]{soltani_detecting_2023}
\bibinfo{author}{Soltani, S.}, \bibinfo{author}{Hosseini~Seno, S.A.},
  \bibinfo{year}{2023}.
\newblock \bibinfo{title}{Detecting the software usage on a compromised system:
  A triage solution for digital forensics}.
\newblock \bibinfo{journal}{Forensic Science International: Digital
  Investigation} \bibinfo{volume}{44}, \bibinfo{pages}{301484}.
\newblock \URLprefix \url{https://doi.org/10.1016/j.fsidi.2022.301484}.
\bibitem[{Tilbury(2023)}]{tilbury_2023_finding}
\bibinfo{author}{Tilbury, C.}, \bibinfo{year}{2023}.
\newblock \bibinfo{title}{Finding evil wmi event consumers with disk
  forensics}.
\newblock \URLprefix
  \url{https://www.sans.org/blog/finding-evil-wmi-event-consumers-with-disk-forensics}.
\bibitem[{Ullrich and Flook(2022)}]{sans_isc_prefetch_2022}
\bibinfo{author}{Ullrich, J.}, \bibinfo{author}{Flook, L.},
  \bibinfo{year}{2022}.
\newblock \bibinfo{title}{Forensic value of prefetch}.
\newblock \bibinfo{howpublished}{SANS Internet Storm Center Diary}.
\newblock \URLprefix \url{https://isc.sans.edu/diary/29168}.
\bibitem[{{Velocidex}(2025)}]{velociraptor_sigma_docs}
\bibinfo{author}{{Velocidex}}, \bibinfo{year}{2025}.
\newblock \bibinfo{title}{Sigma in velociraptor}.
\newblock \bibinfo{howpublished}{Online Documentation}.
\newblock \URLprefix
  \url{https://sigma.velocidex.com/docs/sigma_in_velociraptor/}.
\bibitem[{Velociraptor(2025)}]{velociraptor_2025_sigma}
\bibinfo{author}{Velociraptor}, \bibinfo{year}{2025}.
\newblock \bibinfo{title}{Sigma tracker :: Velociraptor - digging deeper!}
\newblock \URLprefix
  \url{https://docs.velociraptor.app/docs/troubleshooting/debugging/global/vql/plugins/sigma/}.
\bibitem[{{Velociraptor}(2025)}]{velociraptor_overview_nd}
\bibinfo{author}{{Velociraptor}}, \bibinfo{year}{2025}.
\newblock \bibinfo{title}{Velociraptor overview}.
\newblock \bibinfo{howpublished}{Documentation}.
\newblock \URLprefix \url{https://docs.velociraptor.app/docs/overview/}.
\bibitem[{Velociraptor(2025a)}]{velociraptor_2025_windowsforensicsprefetch}
\bibinfo{author}{Velociraptor}, \bibinfo{year}{2025}a.
\newblock \bibinfo{title}{Windows.forensics.prefetch :: Velociraptor - digging
  deeper!}
\newblock \URLprefix
  \url{https://docs.velociraptor.app/artifact_references/pages/windows.forensics.prefetch/}.
\bibitem[{Velociraptor(2025b)}]{velociraptor_2025_windowsforensicsusn}
\bibinfo{author}{Velociraptor}, \bibinfo{year}{2025}b.
\newblock \bibinfo{title}{Windows.forensics.usn}.
\newblock \URLprefix
  \url{https://docs.velociraptor.app/artifact_references/pages/windows.forensics.usn/}.
\bibitem[{{Velociraptor}(2026a)}]{velociraptor_vql_clock}
\bibinfo{author}{{Velociraptor}}, \bibinfo{year}{2026}a.
\newblock \bibinfo{title}{clock :: {VQL} reference}.
\newblock \URLprefix
  \url{https://docs.velociraptor.app/vql_reference/event/clock/}.
\bibitem[{{Velociraptor}(2026b)}]{velociraptor_vql_dedup}
\bibinfo{author}{{Velociraptor}}, \bibinfo{year}{2026}b.
\newblock \bibinfo{title}{dedup :: {VQL} reference}.
\newblock \URLprefix
  \url{https://docs.velociraptor.app/vql_reference/other/dedup/}.
\bibitem[{{Velociraptor}(2026c)}]{velociraptor_vql_foreach}
\bibinfo{author}{{Velociraptor}}, \bibinfo{year}{2026}c.
\newblock \bibinfo{title}{foreach :: {VQL} reference}.
\newblock \URLprefix
  \url{https://docs.velociraptor.app/vql_reference/popular/foreach/}.
\bibitem[{Yang et~al.(2025)Yang, Lin and Chao}]{yang_enhancing_2025}
\bibinfo{author}{Yang, H.C.}, \bibinfo{author}{Lin, I.L.},
  \bibinfo{author}{Chao, Y.H.}, \bibinfo{year}{2025}.
\newblock \bibinfo{title}{Enhancing traditional reactive digital forensics to a
  proactive digital forensics standard operating procedure ({P-DEFSOP}): A case
  study of {DEFSOP} and {ISO} 27035}.
\newblock \bibinfo{journal}{Applied Sciences} \bibinfo{volume}{15},
  \bibinfo{pages}{9922}.
\newblock \URLprefix \url{https://doi.org/10.3390/app15189922}.

\end{thebibliography}
\end{document}